\documentclass[10pt,letterpaper]{article}
\usepackage{opex3}
\usepackage{cite}
\newcommand{\ket}[1]{\left| #1 \right\rangle}
\newcommand{\braket}[2]{\left\langle {#1{\left| \vphantom{#1 #2} \right.} #2} \right\rangle}
\newcommand{\qo}[1]{``#1''}

\usepackage{array}
\newcolumntype{C}{>{\(}c<{\)}}
\newcolumntype{L}{>{\(}l<{\)}}
\renewcommand{\epsilon}{\varepsilon}
\renewcommand{\phi}{\varphi}

\begin{document}

\title{Integrated multi vector vortex beam generator}
\author{Sebastian A. Schulz,$^{1}$  Taras Machula,$^{1}$ Ebrahim Karimi,$^{1,\ast}$\\ and Robert W. Boyd$^{1,2}$}
\address{$1$ Department of Physics, University of Ottawa, 150 Louis Pasteur, Ottawa, Ontario, K1N 6N5 Canada\\
$2$ Institute of Optics, University of Rochester, Rochester, New York, 14627, USA}
\email{ekarimi@uottawa.ca}
\begin{abstract}
A novel method to generate and manipulate vector vortex beams in an integrated, ring resonator based geometry is proposed. We show numerically that a ring resonator, with an appropriate grating, addressed by a vertically displaced access waveguide emits a complex optical field. The emitted beam possesses a specific polarization topology, and consequently a transverse intensity profile and orbital angular momentum. We propose a combination of several concentric ring resonators, addressed with different bus guides, to generate arbitrary orbital angular momentum \textit{qudit} states, which could potentially be used for classical and quantum communications. Finally, we demonstrate numerically that this device works as an orbital angular momentum sorter with an average cross-talk of $-10$dB between different orbital angular momentum channels.
\end{abstract}
\section{Introduction}
It is well known that a light beam carries a linear momentum, which can be transferred to an absorbing or reflecting medium~\cite{allen:03}. This linear momentum of light is proportional to the time-average of the Poynting vector. The Poynting vector associated with a plane wave is purely longitudinal and has no transverse components, consequently the linear momentum is parallel to the propagation direction. However, since a plane wave carries infinite energy it is not physically realizable. Therefore, all generated optical beams are solutions of the paraxial wave equation, leading to a non-vanishing component of the linear momentum in the transverse plane~\cite{siegman:86}. Such a paraxial beam might possess a net orbital angular momentum (OAM) in the propagation direction as a consequence of having a finite transverse linear momentum~\cite{allen:03,frankearnold:08}. Stationary azimuthal linear momentum leads to a quantized OAM of $\ell\hbar$ per photon in the propagation direction, where $\ell$ and $\hbar$ are an integer and the reduced Planck constant respectively~\cite{allen:92,berry:98}. The phase front associated with such a beam forms a helical shape of $\exp{(i\ell\phi)}$, where $\phi$ is the azimuth angle of the polar coordinate system. Additionally, light possesses an independent form of angular momentum, called spin angular momentum (SAM). The SAM is related to the vectorial property of optical fields, with a value of $\pm\hbar$ per photon, corresponding to left and right hand circularly polarized light, respectively~\cite{poynting:09,beth:35}. Although the SAM and OAM are independent features of the optical field, they can interact with each other under certain circumstances~\cite{zhao:07,brasselet:09,marrucci:11}.

Many methods to generate and manipulate optical OAM have been presented in the past, due to the potentially use of OAM in a wide variety of applications, such as optical lithography, astronomy, optofluidics and optical telecommunication~\cite{wang:08,foo:05,he:95,paterson:01,hell:07,gibson:04}. Additionally to the aforementioned classical applications, optical OAM opens up a promising perspective in quantum communication, where a qudit state can be used to encode information~\cite{mair:01,molina:07,barreiro:08,boyd:11}. Implementing a single photon \textit{qudit} state instead of a multi photon \textit{quNit} (N=d) state results in a significantly lower power consumption during state preparation, transmission and detection processes. The previously suggested approaches to generate and manipulate the optical OAM include computer-generated holograms screened on a spatial light modulator (SLM), astigmatic mode converters, spiral phase plates and spin-to-orbit conversion in inhomogeneous birefringent plates ($q$-plates)~\cite{allen:92,bazhenov:92,beijersbergen:94,marrucci:06}. These techniques, however, have several limitations. For instance, beam generation using SLMs has a low efficiency and the beam quality is bounded by the pixel size of nematic liquid crystal cells in the spatial light modulator, while the other approaches are static and cannot be dynamically controlled.

Integrated photonics for quantum computation has generated significant interest, as it allows for a reduced footprint and energy consumption compared to bulk optics components, as well as improved stability~\cite{Politi:08,O'Brien:09}. Additionally, silicon photonics based devices are complementary metal--oxide--semiconductor (CMOS) compatible, offering the opportunity of large scale fabrication and integration with electronic circuits. Y. F. Yu et al. proposed an integrated photonics model to generate optical OAM, using a ring resonator surrounded by a group of nano-rods~\cite{yu:10}. The ring resonator is side coupled to a bus waveguide, with nano-rods at the outer circumference of the ring resonator acting as scatterers for the local evanescent field. Each nano-rod has a specific phase delay with respect to neighboring nano-rods, leading to a helical phase front and therefore a finite OAM. More recently, X. Cai et al. fabricated a compact micro optical vortex beam emitter on a silicon-on-insulator chip, where the nano-rods proposed by Y. F. Yu. et al. were replaced by an in-ring angular grating~\cite{cai:12}. In fact, the beam generated by either angular nano-rods or a  grating does not carry a pure OAM value, instead the emitted light beam owns a space-variant polarization pattern. 
While providing a significant reduction in footprint when compared to other methods for the generation of OAM beams, these methods are not suitable for the generation of OAM \textit{qudit} states, as a single ring resonator will only emit \qo{\textit{vector vortex}} beams with a bounded OAM value for a given wavelength. \newline

In this work, we propose a novel design for the generation of vector vortex beams, where the bus waveguide and the ring resonator are located in different planes~\cite{S.Suzuki:92}. Numerical simulations show that this \textit{out of plane} configuration leads to very good coupling between the bus guide and the ring resonator, comparable to the side-coupled configuration, while the quality of the generated vector vortex beam is not affected by the bus waveguide placement. Furthermore, the proposed design gives the opportunity to build up concentric ring resonators with different radii that can be addressed independently, through individual access waveguides. We show numerically that adjusting both the phase and the amplitude of the bus waveguides input can be used to encode a  \textit{qudit} state.
\begin{figure}[t]
\begin{center}
	\includegraphics[width=8cm]{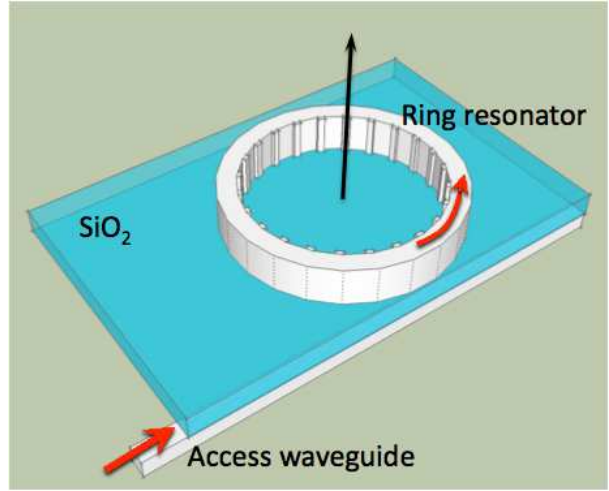}
	\caption{\label{fig:ring_resonator} Schematic of the proposed configuration of a silicon bus wave-guide and ring resonator. The ring resonator is on top of the access waveguide, separated by a $275$ nm thick layer of silica. The angular grating inside the ring resonator is similar to that suggested in Ref.~\cite{cai:12}. }
\end{center}
\end{figure}
\begin{figure*}[t]
\begin{center}
	\includegraphics[width=13cm]{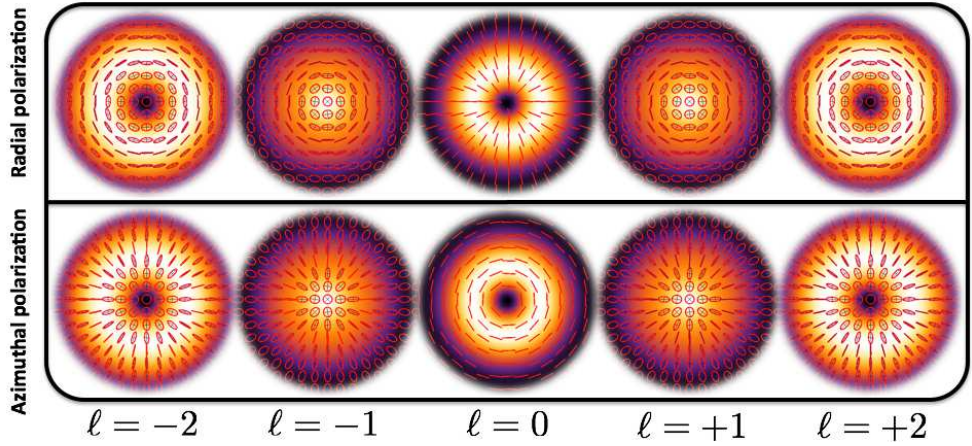}
	\caption{\label{fig:polarizationpattern}Theoretical intensity and polarization patterns for a beam emitted from a ring resonator. The upper row is for light polarized perpendicular to the waveguide, yet in the plane of the resonator, while the lower row is for light polarized parallel to the waveguide, leading to a radial and an azimuthal polarization distribution inside the ring resonator respectively. Each column corresponds to a specific angular phase matching condition and consequently value of OAM. Interestingly, for the case where $\ell=\pm1$, the intensity pattern of the outgoing beam does not form a doughnut shape, and its polarization switches from radial to azimuthal and vice versa. This type of beam is know as the Poincar\'e or polarization-singular beam. In this analysis, we have neglected any possible variations in the radial beam profile due to confinement of the scattering sources and instead a Laguerre-Gauss mode of radial order \textit{zero} was assumed. }
\end{center}
\end{figure*}
\section{Theoretical model}
As discussed in the introduction, it has been shown that a sequence of nano-rods or notches on a waveguide can be used as a grating to scatter the evanescent wave near a ring resonator~\cite{yu:10,cai:12}. Under the correct conditions the emitted beam will carry OAM, with the OAM value being determined by the difference between the resonator mode number and the number of scattering elements. In the previous configurations both the access waveguide and the ring resonator were in the same plane, with a horizontal (side) coupling between them. This horizontal coupling configuration limits the devices to mainly preliminary applications, since the higher dimensions of the OAM Hilbert space stay inaccessible. In order to overcome this issue, we propose a configuration where the waveguide and ring resonator are in a separate plane, as shown in Fig. (\ref{fig:ring_resonator}). In the proposed scheme the access waveguide is still coupled to the ring resonator through evanescent coupling, albeit in the vertical direction. On resonance the light inside the ring resonator satisfies the equation 
\begin{eqnarray}\label{eq:ringresonance}
	 R_0=m\lambda/(2\pi n_{eff}),
\end{eqnarray}
where $m$, $n_{eff}$ and $\lambda$ stand for the ring resonator mode number (a positive integer), the effective index of the waveguide mode and the free-space wavelength, respectively. The ring resonator mode number, $m$, is defined as the number of optical periods in the resonator circumference, $2\pi R_0$. This standing wave possesses a \qo{net} azimuthal linear momentum of $\beta_m=2\pi n_{eff}/\lambda=m/R_0$ in the plane of ring resonator.  
In order to couple this beam out of the ring resonator (into the vertical direction), we choose an appropriate grating with elements on the inside wall of the ring resonator and equal angular spacing, similar to those suggested in Ref.~\cite{cai:12}.

For the remainder of this work we assume that the resonator is on a resonance, with mode number $m$. The grating elements can be designed to be at any relative phase position of the standing wave. For instance, let us assume a ring resonator with $m=1$, i.e.  only one optical period is contained within the ring. Four symmetric objects inside the ring resonator with a relative angle of $\pi/2$ could be used to pick up a beam with a relative phase of $\pi/2$. However, adding additional scattering elements improves the quality of the phase front as well as the OAM state purity, but does not affect the beam structure. Instead, a higher order mode of the ring resonator should be excited in order to change the beam phase-front. To clarify this point, let us consider a ring resonator with $m=4$ and the four symmetric scattering objects from the previous example. In this case, the scattering elements pick up the coupled wave at four different positions, all of which are in phase. Therefore, in the absence of diffraction effects, the scattered beam possesses a uniform phase-font and consequently no OAM. More generally, it has been shown theoretically that the emitted wave from the grating interferes constructively if the following angular phase matching condition is satisfied;
\begin{eqnarray}\label{eq:oammode}
	\ell=m-q,
\end{eqnarray}
where $q$ stands for the number of grating elements~(e.g., see supplementary material of Ref.~\cite{cai:12}).

In addition to the phase property of light, the light inside the ring resonator also has a well defined polarization. The electric field can either lie in the same horizontal plane as the ring resonator, i.e. transverse electric (TE), or orthogonal to this plane, i.e. transverse magnetic (TM) polarization. Note that these definitions of polarization are standard in integrated optics, however they differ from the standard optics definitions. In this work we exclusively use the integrated optics definitions, as given above.

We assume that the grating is made of a linear, isotropic, and nonmagnetic medium. Under this circumstance and assuming a weak scattering regime the {\bf far-field} projection (observed at position $\mathbf{r}$) of the optical fields $(\mathbf{E},\mathbf{B})$ of the light scattered by each grating pitch (located at $\mathbf{r}_0=R_0\widehat{r}$) is given by:
\begin{eqnarray}\label{eq:eandb}
	\mathbf{E}(\mathbf{r})&=&\mathbf{A}(\mathbf{r}_0)\,\frac{e^{ikr}}{r}\cr
	\mathbf{B}(\mathbf{r})&=&\frac{\mathbf{k}}{c}\times\mathbf{A}(\mathbf{r}_0)\,\frac{e^{ikr}}{r},
\end{eqnarray}
where $c$ and $\mathbf{k}=k\,\widehat{\mathbf{k}}$ are the speed of light in vacuum and the scattered beam wave-vector (along $\mathbf{r}-\mathbf{r}_0$), respectively. $\mathbf{A}(\mathbf{r}_0)\propto-\chi\,\widehat{\mathbf{k}}\times\left(\widehat{\mathbf{k}}\times\mathbf{E}(\mathbf{r}_0)\right)$ is the scattering amplitude, where $\chi$ is the grating susceptibility and $\mathbf{E}(\mathbf{r}_0)$ is the evanescent wave at the grating~\cite{born:97}. As we expected, both $\mathbf{E}$ and $\mathbf{B}$ obey the Maxwell equations and therefore $\mathbf{E}$, $\mathbf{B}$ and $\mathbf{k}$ are mutually orthogonal vectors. From this follows intuitively that for the case of a TM mode the light scattered by the grating is in the ring resonator plane and will not contribute to the \textit{out of plane radiation} (in fact, the scattering amplitude $\mathbf{A}(\mathbf{r}_0)$ vanishes). Therefore, for the remainder of this work we will consider the case of TE polarization only, where the electric field inside the ring resonator as well as the evanescent wave are both either radially or azimuthally polarized, corresponding to electric fields orthogonal to or along the propagation direction in the access waveguide. Hence, the scattered \textit{optical} field of Eq.~(\ref{eq:eandb}) is either radially or azimuthally polarized as well, i.e.  $\mathbf{A}(\mathbf{r}_0)\propto A(\mathbf{r}_0)\widehat{r}$ or $\mathbf{A}(\mathbf{r}_0)\propto A(\mathbf{r}_0)\widehat{\phi}$, where $\widehat{r}$ and $\widehat{\phi}$ are the radial and azimuthal unit vectors in the polar coordinates.

Therefore the overall scattered optical field has a complex structure in the far-field. Assuming the angular condition stated in Eq.(\ref{eq:oammode}), its phase-front forms a helix, described by $e^{i\ell\phi}$, and it is either radially or azimuthally polarized, depending on the polarization of light in the access waveguide. Thus, we can express the optical field emerging from the ring resonator in one of the following forms:
\begin{itemize}
\item {radially polarized beam}
\begin{eqnarray}\label{eq:outputr}
	\mathbf{E}_r(\mathbf{r})&=&E_0(r,z)\, e^{i\ell\phi}\,\widehat{r}\cr\cr
			&=&\frac{E_0(r,z)}{\sqrt{2}}\left[\ket{L}_\pi\ket{\ell-1}_o+\ket{R}_\pi\ket{\ell+1}_o\right],
\end{eqnarray}
\item{azimuthally polarized beam}
\begin{eqnarray}\label{eq:outputf}
	\mathbf{E}_\phi(\mathbf{r})&=&E_0(r,z)\, e^{i\ell\phi}\,\widehat{\phi}\cr\cr
			&=&\frac{iE_0(r,z)}{\sqrt{2}}\left[-\ket{L}_\pi\ket{\ell-1}_o+\ket{R}_\pi\ket{\ell+1}_o\right],
\end{eqnarray}
\end{itemize}
\begin{figure}[t]
\begin{center}
	\includegraphics[width=8.6cm]{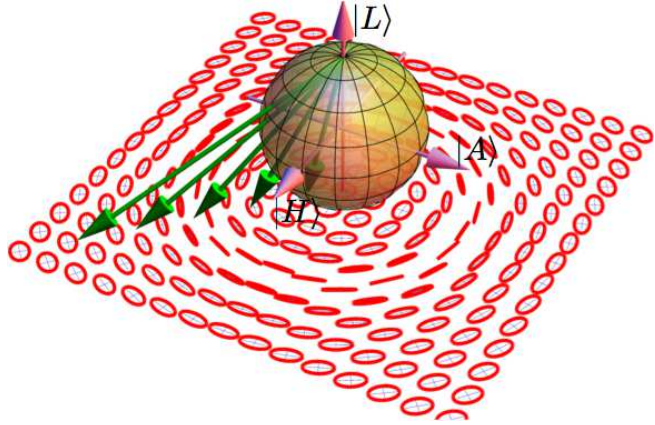}
	\caption{\label{fig:poincare_beam}Sketch showing the stereoscopic mapping of the polarization Poincar\'e sphere on a beam transverse plane for the case of radially polarized beam with $\ell=-1$. The beam possesses a right-handed circular polarization at the center (named C-point), tangent to the sphere's south pole. From this point the polarization transitions through right handed elliptical to linear polarization at the so called L-line, a mapping of the Poincar\'e sphere's equator. The upper hemisphere of the Poincar\'e sphere, however, is mapped to the region outside of the L-line, which therefore has a left handed polarization (initially elliptical, then circular). In the case of a radially polarized beam with $\ell=+1$, the stereoscopic mapping is on the north pole, consequently the polarization handedness changes from right to left handed. }
\end{center}
\end{figure}
where $E_0(r,z)$ is the radial amplitude of the scalar optical field, $\ket{S}_\pi$ and $\ket{O}_o$ are the states of SAM and OAM in the Dirac notation, and $L$ and $R$ stand for left-circularly and right-circularly polarized beams respectively. The polarization and intensity patterns of the output beams for different angular phase matching conditions, i.e. $\ell$, and different polarization coupling conditions are shown in Fig. (\ref{fig:polarizationpattern}). When the angular phase matching is set at $\ell=0$, the emitted beam from the ring resonator does not possess a net OAM value. However, it has either radial or azimuthal polarization, depending on the polarization of light in the access waveguide. These polarization conditions have a singularity at the centre and therefore the beam shape is a doughnut, as can be seen in the middle column of Fig. (\ref{fig:polarizationpattern}). These optical beams have interesting applications in lithography and data storage, where they provide minimum spot size under tight focusing~\cite{dorn:03}. More interestingly, under specific angular phase matching conditions, where $\ell=\pm1$, the emerging beam from the ring resonator forms a complex polarization structure with a non-zero intensity at the beam center, for both types of input polarization. These beams are known as a family of Poincar\'e or polarization-singular beams~\cite{beckley:10, cardano:13, galvez:12}. In fact, there is a \textit{one-to-one} stereoscopic mapping between the polarization Poincar\'e sphere and the beam polarization pattern on the transverse plane. However, for the case of $\ell=-1$, there is a stereoscopic mapping between a \textit{pole-to-pole} path on the polarization Poincar\'e sphere and the beam's transverse plane, as shown in Fig. (\ref{fig:poincare_beam}). At the centre of the beam is a right-handed circular polarization. At a certain region, where the two amplitudes of Eq. (\ref{eq:outputr}) are equal, the polarization has changed to a linear polarization and after this radius, the polarization handedness changes into left-handed circular polarization. The Stokes parameters corresponding to Poincar\'e beams change across the transverse beam plane. In certain regions or points, where the polarization is either a circular ($S_1$ and $S_2=0$) or a linear ($S_3=0$) polarization, the orientation of polarization ellipses or polarization handedness will be undefined. These points or region are called \textit{C-points} or \textit{L-lines}, respectively. %
Furthermore, any paraxial beam as well as the emitted beam from the ring resonator can be expanded in terms of the Laguerre-Gauss modes, since they represent a complete set of solution to the paraxial wave equation. Therefore, and without loss of generality, we choose to truncate this expansion after the first order and have neglected the higher excited radial modes. Moreover, the polarization distribution pattern is altered dramatically by free-space propagation, as has been experimentally verified in \cite{cardano:13}.
\section{Simulation}
In the previous section, we have discussed the properties of a beam emitted from a single ring resonator with scattering elements. The discussion was completely general and no assumption about the geometrical arrangement of the ring resonator and access waveguide was made. However, as stated in the introduction, a single ring resonator cannot be used to generate a qudit state, and we instead propose a system of multiple concentric ring resonators that are addressed through vertically displaced access waveguides. We demonstrate the feasibility and potential applications of this approach through 3D finite difference time domain (FDTD) simulations, using the \textit{Lumerical 8 package}. For the remainder of this paper, we assume that both the ring resonators and the access waveguides are made of $220$ nm high silicon ($500$ nm width) encapsulated in SiO$_{2}$ and the two layers are separated by $275$ nm of silica. This layer separation was chosen as it provides good coupling to a resonance at $1542$ nm, our design wavelength. A top silica layer was included, to provide a comparable structure to those reported in \cite{cai:12}.

\begin{figure}[t]
\begin{center}
	\includegraphics[width=9cm]{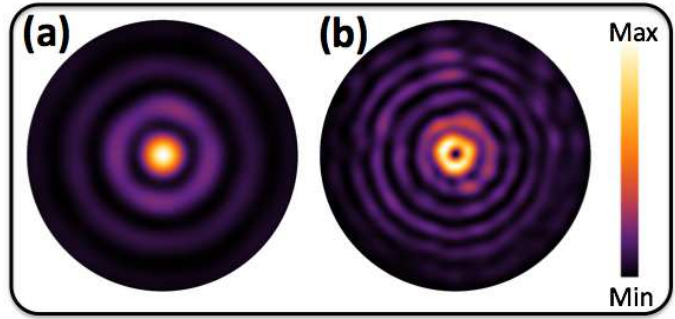}
	\caption{\label{fig:singlering_intensity} Simulated far-field intensity distribution of the beam emitted from a ring resonator, for (a) $\ell=1$ and (b) $\ell=2$. The absence/existence of a doughnut pattern prove the vortex properties of these beam. However, the radial patterns show additional maxima due to angular diffraction for the finite sized source.}
\end{center}
\end{figure}

As a first step, we consider a simple case, where a single ring resonator couples to an access waveguide vertically, demonstrating that the vertical coupling does not affect the generation of OAM beams. This ring resonator has a radius of $R_0=3.9\mu$m and the resonance at $1542$ nm has mode number $m=39$. Two instances of this ring resonator with gratings consisting of $q_1=38$ and $q_2=37$ elements were simulated, giving an angular phase matching (and therefore OAM) of $\ell_1=1$ and $\ell_2=2$, respectively.  Figure \ref{fig:singlering_intensity} shows the simulated intensity patterns of the emitted beams in the far-field zone. The results confirm that the emitted optical field for the case of $\ell_2=2$ possesses a vortex structure, and in comparison to the horizontal coupling demonstrated in \cite{cai:12}, the emitted optical field is not affected by the change to a vertically displaced access waveguide. For the case of  $\ell_1=1$ the emitted beam has a non-zero intensity at the centre, as discussed in the previous section. Apart from higher radial mode excitations, the simulated far-field patterns are in excellent agreement with our theoretical model shown in Fig. (\ref{fig:polarizationpattern}).

While a single ring resonator emits one of the complex polarization structures shown in Figs.  (\ref{fig:polarizationpattern}) and (\ref{fig:singlering_intensity}) this is not yet an OAM qudit state. In order to generate and manipulate OAM qudit states, we must address several concentric ring resonators at the same time. These ring resonators must be coupled to separate access waveguides and must have a resonance at the same wavelength, such that a superposition state can be formed.
\begin{figure}[t]
\begin{center}
	\includegraphics[width=8cm]{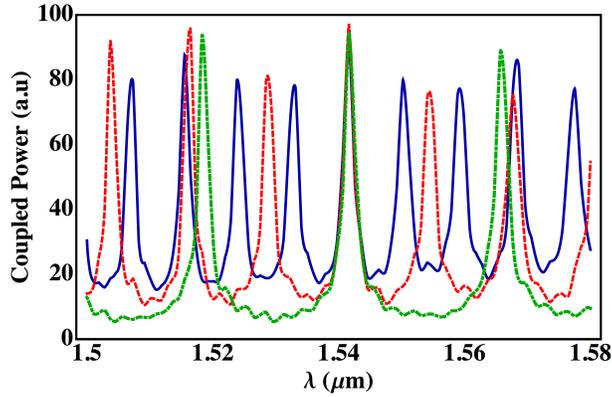}
	\caption{\label{fig:threering_spectrum} Spectrum of three concentric ring resonators: green dot-dashed, red dashed and blue solid lines correspond to inner, middle and outer resonators, respectively. All resonators have a resonance at $1542$ nm, allowing for the formation of a superposition state. The resonator Q factors are between 700 and 1000.}
\end{center}
\end{figure}
Figure \ref{fig:threering_spectrum} shows the resonance spectrum for three concentric ring resonators. The resonators are designed such that they all have a resonance at $1542$ nm. When a single resonator is addressed, we observe a low crosstalk ($~10$ dB) between the resonators. Due to the excessively large simulation time required for the three resonator system, most calculations in this work are performed for a two resonator system. However, the results are general and can be extended to systems with three or more ring resonators.
\section{Applications}
To demonstrate the potential of the proposed system we will now discuss several potential and interesting applications.\newline
\begin{figure*}[t]
\begin{center}
	\includegraphics[width=13cm]{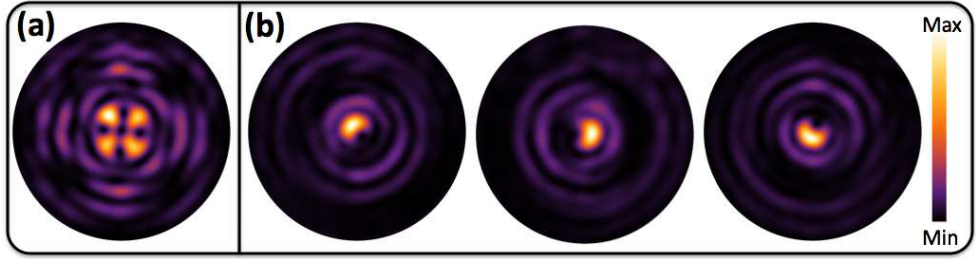}
	\caption{\label{fig:qubitstate} Simulated intensity pattern of a superposition state generated by two concentric ring resonators. (a) The beam is a superposition of $\ell=+2$ and $\ell'=-2$, and (b) $\ell=+2$ and $\ell'=+1$, respectively. In (b) the relative phase of the input signals for the two waveguides is varied, leading to a rotation of the superposition pattern. In (b), the intensity patterns, from left to right, correspond to relative phase of $\chi=0^{\circ}$, $\chi=120^{\circ}$, and $\chi=240^{\circ}$, respectively.}
\end{center}
\end{figure*}
\newline

\subsection{Quantum information}
Implementing a novel secure quantum key distribution (QKD) protocol in the OAM state space requires the availability of higher dimensions. In this case, different OAM bases of a single photon can be used as an individual alphabet to encode the information~\cite{boyd:11}. As shown previously, a ring resonators can be used to emit a variety of different OAM values along a \textit{common} propagation axis, running through the ring's centre, providing access to the higher dimensions of the OAM space. When two or more concentric resonators are simultaneously excited each will emit its own vector vortex beam, independent of the other resonators. Addressing these ring resonators coherently leads to a coherent superposition of the emitted vector vortex beams, i.e.  $\sum_{\ell} c_\ell \ket{\ell}$, since the rings have matched resonances, assuming an identical input polarization. Therefore, without loss of generality, the polarization of the output beam can be factorized out. Let us now consider the simplest case, where the information is encoded onto two \textit{mutually} exclusive bases of OAM. Such a bi-dimensional state is known as a qubit with 
\begin{eqnarray}\label{eq:oamqubit}
	\ket{\psi}=\cos{\left(\frac{\theta}{2}\right)}\ket{\ell}+e^{i\chi}\sin{\left(\frac{\theta}{2}\right)}\ket{\ell'},
\end{eqnarray}
where $\braket{\mathbf r}{\ell}=e^{i\ell\phi}$ is the OAM value, and $\theta$ and $\chi$ are the polar and azimuth angles of the state in the Bloch sphere, which define the qubit's amplitude and a relative phase, respectively.  Analogous to polarization states, the bi-dimensional OAM superposition state can be mapped over an OAM Poincar\'e sphere, where the equator describes an equal superposition state having petal shapes with different relative phases~\cite{padgett:99}.

To demonstrate that our system can be used to generate and control such states, we simulate two different superposition states. The first is a state consisting of equal magnitude yet opposite sign OAM values, more specifically $\ell'=-\ell=2$, while the second has different magnitude yet equal sign OAM values, $\ell=2$ and $\ell'=1$. Figure \ref{fig:qubitstate}(a) shows the first case, where two resonators are excited coherently and the gratings are designed such that the inner ring and outer ring emit a beam with $\ell=+2$ and $\ell'=-2$, respectively. As can be seen from the far field pattern, the emitted beam forms a petal shape similar to a Hermite-Gauss mode of HG$_{1,1}$ (the emerging beam is a superposition of $\exp{(2i\phi)}$ and $\exp{(-2i\phi)}$, i.e. $\cos{2\phi}$ with additional angular diffraction effects). In addition to forming superposition states, the relative phase of the two components of the superposition state, i.e.  $\chi$, can be controlled. Figure \ref{fig:qubitstate}(b) shows the simulated intensity pattern for the second case, where the inner and outer rings emit a beam with $\ell=+2$, and $\ell'=+1$ respectively. For this beam the intensity pattern is known as a \qo{C-beam}. If a phase delay is introduced in the access waveguide of one resonator, then the relative phase between two OAM states in the superposition changes. Therefore, the C-beam intensity pattern rotates, as shown in Fig. \ref{fig:qubitstate}(b).

\subsection{Angular phased array} 
As shown in Fig. \ref{fig:qubitstate}(b), the relative phase of the emitted beams can be adjusted through a control of the relative phase of the input waveguides. Interesting effects also occur if this idea is applied to concentric ring resonators emitting the same vector vortex beam, i.e.  same wavelength, OAM value and polarization. Such rings form an angular array of coherent sources that are separated by a few wavelengths, analogous to a phased antenna array. In this angular phased array, changes in the relative phase of the different emitters lead to an angular displacement of the emission pattern, demonstrated through a change in the relative intensity of the different radial orders. \newline
\begin{figure}[h]
\begin{center}
	\includegraphics[width=8cm]{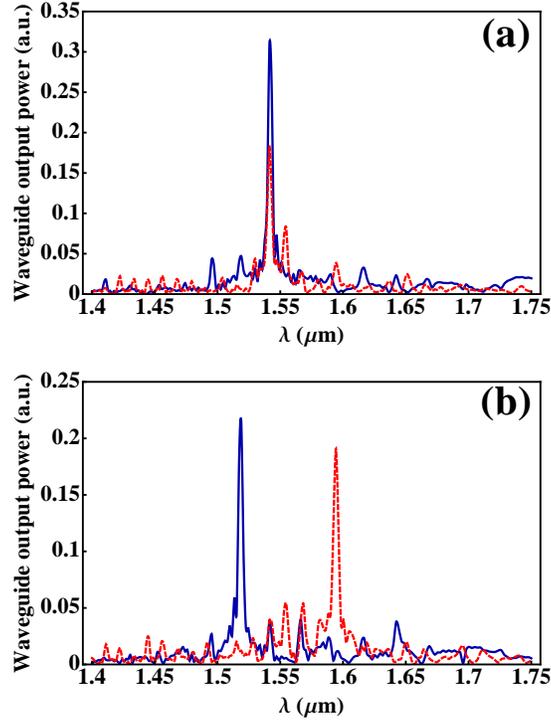}
	\caption{\label{fig:sorter} Optical output power of the access waveguides for ring resonators that are set up for $\ell=+1$ (inner ring) and $\ell=+2$ (outer ring) at $1542$ nm. Red dashed and blue solid curves correspond to the output power from the access waveguide of the inner and outer ring resonators, respectively. (a) Shows the output beam power from access waveguides when illuminated with an equal superposition of $\ell=+1$ and $\ell=+2$. The  power unbalance is due to different out-coupling efficiencies for the outer and inner resonators. We can see a clear peak at the resonance of $1542$ nm. (b) Output power when illuminated with a superposition of $\ell=-2$ and $\ell=+2$ beams. The matched ring resonator (outer) couples at the correct resonance, while the other ring resonator (inner) couples at resonances for which the angular phase matching condition is met. Here we should note that the large bandwidth of the illumination is necessary for a shorter FDTD simulation time, however, experimentally the input could be chosen such that only the resonance of interest (here $1542$ nm) would be excited.}
\end{center}
\end{figure}
%
\subsection{OAM sorter} 
Quantum communication and QKD applications using OAM states require the ability to read out the encoded information and therefore an OAM sorter. In order to provide a scalable, i.e. mass producible system, an integrated, small scale sorter is required. All available sorters have a relatively large singularity size, on the order of a few micrometers, and require bulk optics components, which cannot be used during on chip integration~\cite{leach:02,berkhout:10,karimi:12,sullivan:12}. However, due to the time (and position) reversal invariance of Maxwell's equations, a grating can act as both an emitter from and a coupler to a chip. Therefore, our resonator structures should not only be able to emit beams carrying OAM - they should also couple such beams into the access waveguide of the resonator. This effect was postulated for a single ring resonator in \cite{cai:12}. As shown in Fig. (\ref{fig:sorter}), coupling does indeed occur if the rings are excited by a suitable vortex beam at vertical incidence, assuming that the optical axis is running through the center of the ring resonators. Furthermore, the coupling is selective, i.e.  a ring resonator is only excited if the wavelength and the OAM value of the incident beam match the design of the ring resonator. As such, the OAM of an incoming beam can be measured by simply observing the output of ring resonators of known designs. If several concentric rings are used then a larger alphabet of values can be detected simultaneously and superposition states can be decomposed into their individual OAM components.
As a drawback, it is worth mentioning that there is a non-zero ($-10$dB) crosstalk among different OAM values in the sorter configuration, which compares favorably to other vector vortex sorter (e.g. $-8$ dB for a $q$-plate)~\cite{karimi:09}. A reduction of this crosstalk will require a design optimization, to be addressed in future work.\newline
\section{Conclusions}
We present an approach to generate and manipulate the OAM of light in an on-chip setup, based on multiple concentric ring resonators. It has been shown that the emitted beams from the present configuration, as well as other ring resonator based approaches, form complex patterns of polarization, known as  vector vortex beams. To our knowledge this setup is the first that allows the generation and dynamic control of OAM superposition states in an integrated fashion. Therefore, we believe that this design is a key component in the quest to encode and decode information using the OAM of light.\newline
\section*{Acknowledgments} 
Authors acknowledge the support of the Canada Excellence Research Chairs (CERC) program.

\end{document}